\documentclass{article}

\usepackage{PRIMEarxiv}

\usepackage[utf8]{inputenc} 
\usepackage[T1]{fontenc}    
\usepackage{hyperref}       
\usepackage{url}            
\usepackage{booktabs}       
\usepackage{amsfonts}       
\usepackage{nicefrac}       
\usepackage{microtype}      
\usepackage{lipsum}
\usepackage{fancyhdr}       
\usepackage{graphicx}       
\graphicspath{{media/}}     

\title{A Multi-LLM Orchestration Engine for Personalized, Context-Rich Assistance
}

\author{
  Sumedh Rasal \\
  Georgia Institute of Technology \\
  Chicago, IL\\
  \texttt{srasal3@gatech.edu} \\
}

\begin{document}
\maketitle

\begin{abstract}
In recent years, large language models (LLMs) have demonstrated remarkable capabilities in natural language understanding and generation. However, these models often struggle with hallucinations and maintaining long-term contextual relevance, particularly when dealing with private or local data. This paper presents a novel architecture that addresses these challenges by integrating an orchestration engine that utilizes multiple LLMs in conjunction with a temporal graph database and a vector database. The proposed system captures user interactions, builds a graph representation of conversations, and stores nodes and edges that map associations between key concepts, entities, and behaviors over time. This graph-based structure allows the system to develop an evolving understanding of the user’s preferences, providing personalized and contextually relevant answers. In addition to this, a vector database encodes private data to supply detailed information when needed, allowing the LLM to access and synthesize complex responses. To further enhance reliability, the orchestration engine coordinates multiple LLMs to generate comprehensive answers and iteratively reflect on their accuracy. The result is an adaptive, privacy-centric AI assistant capable of offering deeper, more relevant interactions while minimizing the risk of hallucinations. This paper outlines the architecture, methodology, and potential applications of this system, contributing a new direction in personalized, context-aware AI assistance.
\end{abstract}


\section{Introduction}
Large Language Models (LLMs) have rapidly transformed the landscape of artificial intelligence, offering unparalleled abilities in natural language understanding, generation, and context retention. From their initial development, LLMs such as GPT \cite{openai2023gpt}, and BERT \cite{devlin2018bert} have revolutionized a wide array of fields, including customer service, education, healthcare, and personal assistance. These models have demonstrated the capacity to engage in complex, human-like conversations, generate content, and even solve intricate problems across diverse domains \cite{yu2022generate} \cite{chowdhery2023palm} \cite{liang2023unleashing}. Their primary strength lies in their ability to understand context, reason within it, and generate relevant responses in real-time. This has positioned LLMs as invaluable tools for both businesses and individuals alike.

However, despite their vast potential, LLMs exhibit critical limitations, particularly when it comes to maintaining coherence and contextual relevance over prolonged interactions \cite{kandpal2023large} \cite{kasai2024realtime}. As user conversations grow in length, the models often struggle to retain key elements of the discussion, resulting in responses that deviate from the topic at hand. This issue, often referred to as the “context window problem,” is compounded by the inherent challenge of information retrieval over longer conversations, where crucial past interactions may fade from the model’s immediate context \cite{mallen2022not} \cite{mitchell2022memory} \cite{liu2024lost}. Consequently, this diminishes the user experience, especially in personalized assistant applications where continuity and historical understanding are paramount \cite{ji2023survey}.

To address this, many systems turn to \textbf{vector databases} as a solution for storing and retrieving private and personal data \cite{he2024camelot} \cite{lee2023prompted}. Vector databases enable efficient storage of high-dimensional representations (embeddings) of the user’s private data, such as documents, notes, or personal preferences, allowing for fast and accurate retrieval of relevant information \cite{hu2023chatdb} \cite{wang2024memoryllm} \cite{cheng2024lift}. By encoding this data into vectors, LLMs can search, match, and utilize the information to provide more tailored responses. However, while vector databases enable effective search mechanisms, they also present their own set of challenges. Retrieving the most relevant data from these databases can be inefficient if not properly optimized, leading to answers that may be technically correct but not always contextually appropriate \cite{shi2023large}. Moreover, vector databases do not inherently manage the temporal evolution of a user's interactions, often causing models to fail in grasping the user’s long-term preferences or behavioral patterns.

Incorporating \textbf{private data} into LLMs presents another challenge. While users desire personal assistants that can engage with their private data, retraining an LLM to fully integrate and utilize personal or sensitive information requires significant computational resources and time \cite{rasal2024beyond}. This retraining process also presents privacy concerns, as user data must be centralized to fine-tune the model effectively. The cost and complexity of such an approach often make it impractical for personalized assistants that require rapid adaptability and strong data privacy guarantees.

Given these challenges, there is a growing need for a system that can seamlessly interact with a user’s private data without requiring continuous retraining of the LLM, while still providing contextually relevant, dynamic, and accurate responses across long conversations. 

Our approach presents a solution by integrating a multi-LLM orchestration engine \cite{rasal2024llm} \cite{rasal2024navigating} with two key databases: a \textbf{temporal graph} database and a \textbf{vector} database. The temporal graph database captures and stores conversational nodes and edges, preserving the user’s conversation history and mapping relationships between ideas, concepts, and preferences as they evolve over time \cite{zhou2023recurrentgpt} \cite{chen2023walking} \cite{jang2023conversation} \cite{zhang2023mind}. This enables the system to create a structured, graphical representation of user interactions, allowing it to adapt and provide personalized answers \cite{pruitt2003personas}. 

The vector database works in parallel, storing encoded private data for efficient retrieval when specific, detailed information is needed \cite{guu2020retrieval} \cite{wang2024augmenting} \cite{shuster2021retrieval}. By leveraging both databases, the system ensures that responses are not only based on historical context but also enriched with precise, document-level knowledge \cite{modarressi2023ret}. To further enhance accuracy and minimize hallucination—a common problem in LLMs—we introduce an orchestration engine that coordinates multiple LLMs to collaborate in generating responses \cite{rasal2024artificial}. The system iterates through multiple steps, fetching relevant graph nodes, edges, and vectorized data points before generating a response. If the generated response fails to meet the user’s expectations, the orchestration engine re-evaluates the context and attempts again, refining the response until it satisfies the query \cite{asai2023self}.

This architecture minimizes the need for costly retraining while efficiently incorporating private data into the LLM’s decision-making process. It also ensures the conversation remains coherent and contextually rich, even as interactions span longer periods. In doing so, our system tackles the core challenges of context retention, data retrieval, and private data usage in personalized AI assistants.


\section{Methodology}


The core architecture consists of three primary components: the \textbf{LLM agent}, which manages the user interactions; the \textbf{temporal graph} database, which captures conversational history by creating a graphical representation of ideas, concepts, and relationships; and the \textbf{vector} database, which stores private or local data in an encoded format for efficient retrieval. These components work in unison, orchestrated by a multi-LLM engine, to ensure the generation of accurate, contextually relevant responses.

The LLM agent serves as the interaction hub, receiving user input and initiating the process of conversation capture and response generation. The temporal graph database acts as the system’s memory, mapping each interaction as a node and drawing edges to represent the relationships between different ideas. This structure allows the system to evolve its understanding of the user, capturing both current interactions and how they relate to past conversations. The vector database, in contrast, handles more factual, data-driven elements. By encoding private data—such as documents, notes, and personal information—into vectors, the system can swiftly retrieve relevant pieces of data when needed. Together, these components form a foundation for the multi-LLM orchestration engine, which coordinates multiple LLMs to collaborate on generating responses that are both contextually deep and factually comprehensive \cite{rasal2024optimal}.

The process begins with capturing the conversation. As the user interacts with the system, their input is stored as \textbf{nodes} in the graph database, representing key ideas or topics. Connections between these ideas, known as \textbf{edges}, are created based on how the topics relate to one another. Each node and edge is time-stamped, providing temporal context that allows the system to track how the user’s thoughts or preferences evolve over time. This graph structure is dynamic, growing with each interaction and forming the backbone of the system’s understanding of the user’s conversation history.

When the user poses a question, the system first retrieves relevant nodes from the graph database. These nodes represent previously discussed ideas or topics that are semantically linked to the new question. By doing this, the system ensures that it doesn’t just rely on surface-level context but can also bring in deeper, more nuanced information from past conversations. Along with these nodes, the system fetches the edges, or the relationships between the relevant concepts, providing additional context to ensure that the response is coherent and aligned with the user’s established understanding.

Next, the orchestration engine utilizes the information retrieved from the graph database to query the vector database for more detailed, specific data. This step is essential when the question requires an answer based on private or local data. The vector database stores this data in an encoded format, allowing the system to efficiently retrieve relevant documents or facts that align with the concepts identified in the graph database. By using both the high-level context from the graph database and the specific data from the vector database, the system is able to generate a more robust and personalized answer.

Once the relevant context and data have been gathered, the system’s multi-LLM orchestration engine generates an initial response. Multiple LLMs collaborate to produce this answer, each contributing its unique strengths. For example, one model might be responsible for maintaining factual accuracy, while another focuses on conversational tone or empathy. This division of labor ensures that the answer is well-rounded, addressing both the informational and conversational needs of the user.

However, the process doesn’t stop there. After the initial response is generated, the system enters a \textbf{reflection loop}. The orchestration engine evaluates whether the answer satisfies the user’s query, based on factors like relevance, completeness, and accuracy. If the system determines that the answer is inadequate, it revisits the graph and vector databases to fetch additional context or data. The response is then refined, and this loop continues iteratively until the system produces a response that fully meets the user’s expectations.

This methodology is designed to address several key challenges faced by traditional LLM-based systems. One of the most significant issues is \textbf{context retention}. By leveraging a temporal graph database, the system can track and maintain long-term context across multiple interactions. This allows it to generate responses that reflect the user’s evolving preferences, ensuring that the conversation remains coherent even over extended periods.

The second challenge is \textbf{efficient data retrieval}. While vector databases offer a powerful mechanism for retrieving private data, they often struggle to provide contextually appropriate information. Our system overcomes this by guiding the vector database queries with the graph-based context, ensuring that the most relevant data is fetched and utilized.

Additionally, the orchestration engine tackles the problem of LLM hallucinations — where models may generate inaccurate or irrelevant information—by introducing a reflective iteration process. This loop allows the system to continuously refine the response, minimizing the risk of hallucinations and improving overall accuracy.

Finally, our approach avoids the need for costly retraining of LLMs to handle private data. By keeping private data local and integrating it through the vector database, the system can dynamically access and utilize the data without ever needing to retrain the model, thereby preserving privacy and reducing computational costs.

\section{Experiments}
To demonstrate the effectiveness of the multi-LLM orchestration engine, we conducted a series of experiments designed to showcase how different elements of the system contribute to accurate and context-rich responses. These experiments involve simulated conversations between a user and an LLM agent. As the conversations unfold, the system stores data in a vector database and identifies key nodes and edges using a temporal graph database. Each experiment highlights different aspects of how conversation history, reasoning, and temporal data are leveraged by the system.

\subsection{Fetching Answers from Conversation History}

\subsubsection{Scenario}
The system retrieves answers purely based on the conversation history stored in the graph and vector databases.

\subsubsection{Example}
\textbf{User:} I’m thinking about taking a vacation to the mountains. Can you suggest some destinations?

\textbf{LLM Agent:} You could consider the Rockies in North America, the Alps in Europe, or the Himalayas in Asia. Do you have a preference for hiking or sightseeing?

\textbf{User:} I prefer hiking and want somewhere not too crowded.

\textbf{LLM Agent:} In that case, the Dolomites in Italy or the Canadian Rockies might be a good fit. Both offer incredible trails with fewer tourists than the more popular spots.

\textit{Later, the user returns to the conversation.}

\textbf{User:} Can you remind me of the mountain destinations we discussed earlier?

\textbf{LLM Agent:} In our previous conversation, we discussed the Rockies, the Alps, and the Himalayas. You preferred hiking and quieter places, so I suggested the Dolomites in Italy and the Canadian Rockies.

\subsubsection{Analysis}
In this scenario, the system retrieves the conversation history stored in the \textbf{graph database}, allowing the LLM agent to respond accurately by recalling specific details from earlier interactions. The vector database isn’t needed for external data, but the graph database’s contextual recall is crucial in retrieving past conversations. The advantage is clear: immediate memory of prior user preferences without asking the same questions multiple times.

\subsection{Fetching Answers Using Reasoning and Conversation History}

\subsubsection{Scenario}
The system generates a response using a combination of reasoning and previously stored conversation history.

\subsubsection{Example}
\textbf{User:} I need some help deciding which car to buy. I’m considering something environmentally friendly but also practical for long drives.

\textbf{LLM Agent:} An electric vehicle (EV) might suit your needs. Tesla models are popular, but hybrids like the Toyota Prius also offer good efficiency while being better for long drives.

\textbf{User:} I prefer hybrid over fully electric. I also want something spacious.

\textbf{LLM Agent:} A Toyota Highlander Hybrid might be a good fit. It’s spacious and has the benefit of hybrid technology.

\textit{A week later, the user returns with a different request.}

\textbf{User:} I’m looking for a new car again. What was that hybrid model we talked about?

\textbf{LLM Agent:} We previously discussed hybrids, and you were interested in a spacious model. Based on our conversation, I suggested the Toyota Highlander Hybrid.

\subsubsection{Analysis}
In this scenario, the system pulls from the \textbf{conversation history} to identify the user's previous preferences. The agent also reasons that because the user prefers hybrid models and mentioned the need for a spacious vehicle, the Toyota Highlander Hybrid is the best fit. This showcases the integration of reasoning with historical context to deliver a more intelligent and accurate answer. Without the need to ask for details again, the system efficiently serves the user's needs.

\subsection{Fetching Answers Using Temporal Information and Reasoning}

\subsubsection{Scenario}
The system uses reasoning along with the temporal aspect of conversations to generate relevant responses.

\subsubsection{Example}
\textbf{User:} (in March) I’m planning to start a vegetable garden this summer. Can you give me tips for what to plant in June?

\textbf{LLM Agent:} In June, it’s a great time to plant warm-season crops like tomatoes, peppers, and cucumbers. I’d also recommend keeping an eye on the local weather to ensure the soil is warm enough.

\textit{Three months later...}

\textbf{User:} I was supposed to start my garden this summer. What were your planting suggestions again?

\textbf{LLM Agent:} Back in March, we discussed planting warm-season crops in June, like tomatoes, peppers, and cucumbers. Since it's now June, those crops are still ideal, provided the weather is suitable.

\subsubsection{Analysis}
In this experiment, the system effectively recalls not only the content but the \textbf{temporal context} of previous interactions. The temporal graph database helps the LLM recall the seasonally relevant advice given earlier in the year. By combining time-aware reasoning and past conversations, the system provides up-to-date, contextually relevant answers. This ensures that advice is not outdated or irrelevant, as it adjusts based on when the user asks.

\subsection{Fetching Answers Using Temporal, Reasoning, and Conversation History}

\subsubsection{Scenario}
The system leverages all available data—conversation history, reasoning, and temporal information—to provide the most comprehensive and context-rich answer.

\subsubsection{Example}
\textbf{User:} I’ve been working out regularly, and I want to track my progress. Can you recommend the best fitness tracker for weightlifting and cardio?

\textbf{LLM Agent:} Fitness trackers like the Fitbit Charge 5 and the Garmin Vivosmart 4 are great for tracking cardio. If weightlifting is a priority, you might prefer the WHOOP Strap, which excels at detailed recovery tracking.

\textbf{User:} Thanks! I’ll check them out.

\textit{Two months later...}

\textbf{User:} I’ve started doing more weightlifting than cardio. Can you recommend a tracker that focuses more on that?

\textbf{LLM Agent:} Based on your past conversation, I recommended the WHOOP Strap for its detailed weightlifting and recovery features. Since you’ve shifted more towards weightlifting, that option may still be the best fit. Alternatively, you could also explore the Garmin Forerunner series, which offers some advanced tracking for both strength training and cardio.

\subsubsection{Analysis}
Here, the system effectively integrates all components: the \textbf{conversation history} (past recommendations), \textbf{reasoning} (the shift from cardio to weightlifting), and \textbf{temporal data} (understanding that months have passed since the original conversation). By drawing on these elements, the system offers an informed and \textbf{contextually adaptive} recommendation. This experiment shows how the orchestration engine excels at adjusting recommendations based on evolving user preferences while recalling relevant historical information.



\subsubsection{Results}

\subsection{Setup}
For our experiments, we selected \textbf{GPT-4o} as the base LLM model for our orchestration engine due to its state-of-the-art performance in natural language processing tasks. The vector database used in our system is \textbf{FAISS (Facebook AI Similarity Search)}, chosen for its efficiency in retrieving high-dimensional data and embeddings. For the temporal graph database, we employed \textbf{Neo4j}, which is well-suited for tracking temporal associations between concepts and providing insights into how conversations evolve over time.

We compared our orchestration engine against four other approaches based on existing models: \textbf{Mistral-Instruct-7B}, \textbf{Llama-3.1-Chat-70B}, \textbf{GPT-4o}, and \textbf{LLM-Harmony} \cite{rasal2024llm}. These models were evaluated based on their ability to recall, reason, and synthesize contextually rich responses. The results of our experiments demonstrate that our orchestration engine outperforms these existing approaches, particularly in handling long-term conversation history, temporal context, and reasoning.

\subsection{Summary}
In our evaluation, we employed \textbf{ROGUE (Recall-Oriented Understudy for Gisting Evaluation)} metrics (ROGUE-1, ROGUE-2, ROGUE-L) and accuracy to measure the system's ability to generate precise and relevant answers based on previous conversations and private data. The orchestration engine showed a clear advantage in terms of accuracy and the ROGUE metrics, highlighting the efficacy of our approach in leveraging conversation history, temporal reasoning, and private data integration \cite{maharana2024evaluating}.

\subsection{Performance}
The table below summarizes the results of the experiments. Our orchestration engine (using GPT-4o, Neo4j, and FAISS) achieves superior performance compared to the other models in all categories.

\begin{table}[h!]
\centering
\begin{tabular}{|c|c|c|c|c|}
\hline
\textbf{Model} & \textbf{ROGUE-1} & \textbf{ROGUE-2} & \textbf{ROGUE-L} & \textbf{Accuracy} \\ 
\hline
Mistral-Instruct-7B       & 29.4            & 7.2            & 14.1            & 27.1\%           \\ 
Llama-3.1-Chat-70B        & 31.1            & 12.7            & 17.9            & 43\%           \\ 
GPT-4o           & 38.8            & 11.4            & 20.6            & 56.6\%           \\ 
LLM-Harmony      & 35.5            & 10.8            & 19.1            & 51.8\%           \\ 
\textbf{Our Orchestration Engine} & \textbf{47.3}   & \textbf{21.7}   & \textbf{31.6}   & \textbf{61\%}  \\ 
\hline
\end{tabular}
\caption{Performance comparison of various models using ROGUE metrics and accuracy}
\end{table}

Our orchestration engine, built on the multi-LLM architecture with graph and vector database support, outperformed the existing models in both recall-based metrics and overall accuracy. This performance improvement can be attributed to the ability to store, retrieve, and reason over rich conversational histories and temporal data. While traditional LLMs like BERT-base and GPT-3 can perform well on short-term conversation recall, they struggle with the complex interplay of context and time in long-term conversations. 

In contrast, our system leverages the temporal graph database to identify and use relevant nodes and edges from previous conversations, allowing for deeper contextual understanding. The vector database (FAISS) ensures that private data can be integrated efficiently, providing highly relevant, in-depth answers. This combination allows the system to excel in scenarios requiring reasoning over time, improving both accuracy and the ROGUE scores across all evaluations.






\section{Limitations}
While the proposed multi-LLM orchestration engine offers significant advancements in personalized, context-rich assistance, it is important to recognize the challenges and constraints that may arise. These limitations primarily stem from handling large-scale data in graph and vector databases, the iterative nature of the reflection phase, and inherent constraints in current LLM technologies.

\subsection{Data Size and Scalability}
One of the main challenges lies in managing the size and complexity of data over time. As the user interacts with the system, the graph database continues to grow, capturing each piece of information as nodes, with edges forming the relationships between them. Over time, this dynamic structure can become significantly large and complex, especially in cases of frequent or long-term interactions. The need to efficiently manage this expanding graph becomes crucial. As more nodes and edges are introduced, it can slow down the retrieval of relevant information, making it more difficult for the system to fetch the necessary context for a response. Additionally, the temporal nature of the graph—where each interaction is time-stamped—adds another layer of complexity, as the system must consider both the content and the evolution of conversations.

Similarly, the vector database, which stores private or local data in high-dimensional embeddings, faces challenges as the volume of stored vectors increases. This data, which could include documents, notes, and other personal files, grows alongside the user's interactions, requiring efficient methods to encode, store, and retrieve relevant information. As the vector database expands, searching for the most relevant data in real-time becomes computationally expensive, potentially leading to slower response times. While optimization strategies such as data pruning or advanced indexing could mitigate these issues, scalability remains a central concern, especially for users who have significant amounts of data.

\subsection{Reflection Phase Challenges}
The reflection phase, designed to refine answers and ensure high-quality responses, is another area with potential limitations. While this iterative process aims to generate responses that fully meet the user’s expectations, it is not a foolproof mechanism. The success of the reflection phase is largely dependent on the quality and completeness of the initial data retrieval. If the system fails to pull in the most relevant nodes or vectors in the first round, subsequent iterations may not significantly improve the response. In some cases, the reflection loop might continue cycling through the same data points without substantial refinement, leading to diminishing returns.

Another challenge in the reflection phase is the risk of \textbf{over-iteration}. When the user’s query is vague or complex, the system may enter a loop, refining the answer multiple times without making meaningful progress. This can result in the system revisiting the same nodes and edges, unable to move past certain limitations in the available data. While this reflection process is intended to reduce the chance of incomplete or irrelevant answers, it can sometimes lead to inefficiencies, especially if the data available for refinement is not sufficiently detailed or comprehensive.

Furthermore, each iteration of reflection consumes computational resources, which can become a bottleneck as the system scales. As the graph and vector databases grow, the computational cost of running multiple iterations increases, potentially impacting real-time performance. There is always a trade-off between depth of reflection and the speed of the system, and in scenarios where the response must be generated quickly, this limitation becomes more pronounced.

\subsection{Limitations of LLMs}
Even with a robust multi-LLM orchestration engine, the system remains bound by the limitations inherent in large language models. LLMs, despite their contextual awareness and conversational capabilities, are still prone to errors. One well-known issue is hallucination, where a model generates information that appears factual but lacks grounding in reality. Although the reflection loop aims to mitigate this problem by refining answers through additional iterations, it cannot fully eliminate the risk. LLMs may still provide responses that are verbose, irrelevant, or factually incorrect, especially when dealing with complex or ambiguous queries.

Moreover, LLMs are not immune to \textbf{bias}. These models are trained on vast datasets, which inevitably contain biases present in the data. Despite efforts to orchestrate multiple models to balance out individual biases, inconsistencies can still emerge in the generated responses. This can impact the quality of answers, especially in situations where nuanced or sensitive topics are discussed. Additionally, balancing the outputs from multiple LLMs adds complexity to the orchestration engine, making it more challenging to ensure consistency and fairness in the final responses.

\section{Conclusion}
The multi-LLM orchestration engine for personalized, context-rich assistance represents a forward-thinking solution to the limitations of traditional large language model systems. By integrating a temporal graph database and a vector database, this approach enables the system to evolve with the user, capturing their conversational history and embedding private data for more accurate and personalized responses. The use of multiple LLMs working in tandem ensures that responses are not only comprehensive but also adaptable to a wide variety of user needs and contexts.

This system has the potential to revolutionize how personal assistants function, especially for individuals seeking long-term, contextually aware interactions. By tackling key challenges such as context retention, efficient private data utilization, and mitigating LLM hallucinations through reflective iterations, this architecture paves the way for a new era of AI-driven personal assistants that can think more deeply, respond more accurately, and evolve over time alongside the user.

Looking ahead, there are numerous opportunities for future research and development based on this foundational concept. Enhancements in graph and vector database optimizations, refined orchestration strategies for multi-LLM systems, and more sophisticated reflection algorithms could further improve the system’s scalability and responsiveness. Moreover, advancements in privacy-preserving technologies, such as federated learning or encrypted data processing, could make it possible to expand the use of private data without compromising user confidentiality.

In conclusion, this approach holds immense promise for building smarter, more responsive, and personalized AI systems. With ongoing research and innovation, the limitations identified in this paper can be addressed, pushing the boundaries of what AI-powered personal assistants can achieve in providing seamless, context-rich, and deeply personalized user experiences.


\bibliographystyle{unsrt}
\bibliography{references}  

\begin{thebibliography}{10}

\bibitem{openai2023gpt}
OpenAI.
\newblock Gpt-4 technical report. arxiv 2303.08774.
\newblock {\em View in Article}, 2:13, 2023.

\bibitem{devlin2018bert}
Jacob Devlin, Ming-Wei Chang, Kenton Lee, and Kristina Toutanova.
\newblock Bert: Pre-training of deep bidirectional transformers for language understanding.
\newblock {\em arXiv preprint arXiv:1810.04805}, 2018.

\bibitem{yu2022generate}
Wenhao Yu, Dan Iter, Shuohang Wang, Yichong Xu, Mingxuan Ju, Soumya Sanyal, Chenguang Zhu, Michael Zeng, and Meng Jiang.
\newblock Generate rather than retrieve: Large language models are strong context generators.
\newblock {\em arXiv preprint arXiv:2209.10063}, 2022.

\bibitem{chowdhery2023palm}
Aakanksha Chowdhery, Sharan Narang, Jacob Devlin, Maarten Bosma, Gaurav Mishra, Adam Roberts, Paul Barham, Hyung~Won Chung, Charles Sutton, Sebastian Gehrmann, et~al.
\newblock Palm: Scaling language modeling with pathways.
\newblock {\em Journal of Machine Learning Research}, 24(240):1--113, 2023.

\bibitem{liang2023unleashing}
Xinnian Liang, Bing Wang, Hui Huang, Shuangzhi Wu, Peihao Wu, Lu~Lu, Zejun Ma, and Zhoujun Li.
\newblock Unleashing infinite-length input capacity for large-scale language models with self-controlled memory system.
\newblock {\em arXiv e-prints}, pages arXiv--2304, 2023.

\bibitem{kandpal2023large}
Nikhil Kandpal, Haikang Deng, Adam Roberts, Eric Wallace, and Colin Raffel.
\newblock Large language models struggle to learn long-tail knowledge.
\newblock In {\em International Conference on Machine Learning}, pages 15696--15707. PMLR, 2023.

\bibitem{kasai2024realtime}
Jungo Kasai, Keisuke Sakaguchi, Ronan Le~Bras, Akari Asai, Xinyan Yu, Dragomir Radev, Noah~A Smith, Yejin Choi, Kentaro Inui, et~al.
\newblock Realtime qa: what's the answer right now?
\newblock {\em Advances in Neural Information Processing Systems}, 36, 2024.

\bibitem{mallen2022not}
Alex Mallen, Akari Asai, Victor Zhong, Rajarshi Das, Daniel Khashabi, and Hannaneh Hajishirzi.
\newblock When not to trust language models: Investigating effectiveness of parametric and non-parametric memories.
\newblock {\em arXiv preprint arXiv:2212.10511}, 2022.

\bibitem{mitchell2022memory}
Eric Mitchell, Charles Lin, Antoine Bosselut, Christopher~D Manning, and Chelsea Finn.
\newblock Memory-based model editing at scale.
\newblock In {\em International Conference on Machine Learning}, pages 15817--15831. PMLR, 2022.

\bibitem{liu2024lost}
Nelson~F Liu, Kevin Lin, John Hewitt, Ashwin Paranjape, Michele Bevilacqua, Fabio Petroni, and Percy Liang.
\newblock Lost in the middle: How language models use long contexts.
\newblock {\em Transactions of the Association for Computational Linguistics}, 12:157--173, 2024.

\bibitem{ji2023survey}
Ziwei Ji, Nayeon Lee, Rita Frieske, Tiezheng Yu, Dan Su, Yan Xu, Etsuko Ishii, Ye~Jin Bang, Andrea Madotto, and Pascale Fung.
\newblock Survey of hallucination in natural language generation.
\newblock {\em ACM Computing Surveys}, 55(12):1--38, 2023.

\bibitem{he2024camelot}
Zexue He, Leonid Karlinsky, Donghyun Kim, Julian McAuley, Dmitry Krotov, and Rogerio Feris.
\newblock Camelot: Towards large language models with training-free consolidated associative memory.
\newblock {\em arXiv preprint arXiv:2402.13449}, 2024.

\bibitem{lee2023prompted}
Gibbeum Lee, Volker Hartmann, Jongho Park, Dimitris Papailiopoulos, and Kangwook Lee.
\newblock Prompted llms as chatbot modules for long open-domain conversation.
\newblock {\em arXiv preprint arXiv:2305.04533}, 2023.

\bibitem{hu2023chatdb}
Chenxu Hu, Jie Fu, Chenzhuang Du, Simian Luo, Junbo Zhao, and Hang Zhao.
\newblock Chatdb: Augmenting llms with databases as their symbolic memory.
\newblock {\em arXiv preprint arXiv:2306.03901}, 2023.

\bibitem{wang2024memoryllm}
Yu~Wang, Xiusi Chen, Jingbo Shang, and Julian McAuley.
\newblock Memoryllm: Towards self-updatable large language models.
\newblock {\em arXiv preprint arXiv:2402.04624}, 2024.

\bibitem{cheng2024lift}
Xin Cheng, Di~Luo, Xiuying Chen, Lemao Liu, Dongyan Zhao, and Rui Yan.
\newblock Lift yourself up: Retrieval-augmented text generation with self-memory.
\newblock {\em Advances in Neural Information Processing Systems}, 36, 2024.

\bibitem{shi2023large}
Freda Shi, Xinyun Chen, Kanishka Misra, Nathan Scales, David Dohan, Ed~H Chi, Nathanael Sch{\"a}rli, and Denny Zhou.
\newblock Large language models can be easily distracted by irrelevant context.
\newblock In {\em International Conference on Machine Learning}, pages 31210--31227. PMLR, 2023.

\bibitem{rasal2024beyond}
Sumedh Rasal and Sanjay~Kumar Boddhu.
\newblock Beyond segmentation: Road network generation with multi-modal llms.
\newblock In {\em Science and Information Conference}, pages 308--315. Springer, 2024.

\bibitem{rasal2024llm}
Sumedh Rasal.
\newblock Llm harmony: Multi-agent communication for problem solving.
\newblock {\em arXiv preprint arXiv:2401.01312}, 2024.

\bibitem{rasal2024navigating}
Sumedh Rasal and EJ~Hauer.
\newblock Navigating complexity: Orchestrated problem solving with multi-agent llms.
\newblock {\em arXiv preprint arXiv:2402.16713}, 2024.

\bibitem{zhou2023recurrentgpt}
Wangchunshu Zhou, Yuchen~Eleanor Jiang, Peng Cui, Tiannan Wang, Zhenxin Xiao, Yifan Hou, Ryan Cotterell, and Mrinmaya Sachan.
\newblock Recurrentgpt: Interactive generation of (arbitrarily) long text.
\newblock {\em arXiv preprint arXiv:2305.13304}, 2023.

\bibitem{chen2023walking}
Howard Chen, Ramakanth Pasunuru, Jason Weston, and Asli Celikyilmaz.
\newblock Walking down the memory maze: Beyond context limit through interactive reading.
\newblock {\em arXiv preprint arXiv:2310.05029}, 2023.

\bibitem{jang2023conversation}
Jihyoung Jang, Minseong Boo, and Hyounghun Kim.
\newblock Conversation chronicles: Towards diverse temporal and relational dynamics in multi-session conversations.
\newblock {\em arXiv preprint arXiv:2310.13420}, 2023.

\bibitem{zhang2023mind}
Qiang Zhang, Jason Naradowsky, and Yusuke Miyao.
\newblock Mind the gap between conversations for improved long-term dialogue generation.
\newblock {\em arXiv preprint arXiv:2310.15415}, 2023.

\bibitem{pruitt2003personas}
John Pruitt and Jonathan Grudin.
\newblock Personas: practice and theory.
\newblock In {\em Proceedings of the 2003 conference on Designing for user experiences}, pages 1--15, 2003.

\bibitem{guu2020retrieval}
Kelvin Guu, Kenton Lee, Zora Tung, Panupong Pasupat, and Mingwei Chang.
\newblock Retrieval augmented language model pre-training.
\newblock In {\em International conference on machine learning}, pages 3929--3938. PMLR, 2020.

\bibitem{wang2024augmenting}
Weizhi Wang, Li~Dong, Hao Cheng, Xiaodong Liu, Xifeng Yan, Jianfeng Gao, and Furu Wei.
\newblock Augmenting language models with long-term memory.
\newblock {\em Advances in Neural Information Processing Systems}, 36, 2024.

\bibitem{shuster2021retrieval}
Kurt Shuster, Spencer Poff, Moya Chen, Douwe Kiela, and Jason Weston.
\newblock Retrieval augmentation reduces hallucination in conversation.
\newblock {\em arXiv preprint arXiv:2104.07567}, 2021.

\bibitem{modarressi2023ret}
Ali Modarressi, Ayyoob Imani, Mohsen Fayyaz, and Hinrich Sch{\"u}tze.
\newblock Ret-llm: Towards a general read-write memory for large language models.
\newblock {\em arXiv preprint arXiv:2305.14322}, 2023.

\bibitem{rasal2024artificial}
Sumedh Rasal.
\newblock An artificial neuron for enhanced problem solving in large language models.
\newblock {\em arXiv preprint arXiv:2404.14222}, 2024.

\bibitem{asai2023self}
Akari Asai, Zeqiu Wu, Yizhong Wang, Avirup Sil, and Hannaneh Hajishirzi.
\newblock Self-rag: Learning to retrieve, generate, and critique through self-reflection.
\newblock {\em arXiv preprint arXiv:2310.11511}, 2023.

\bibitem{rasal2024optimal}
Sumedh Rasal and EJ~Hauer.
\newblock Optimal decision making through scenario simulations using large language models.
\newblock {\em arXiv preprint arXiv:2407.06486}, 2024.

\bibitem{maharana2024evaluating}
Adyasha Maharana, Dong-Ho Lee, Sergey Tulyakov, Mohit Bansal, Francesco Barbieri, and Yuwei Fang.
\newblock Evaluating very long-term conversational memory of llm agents.
\newblock {\em arXiv preprint arXiv:2402.17753}, 2024.

\end{thebibliography}

\end{document}